\documentstyle[amssymb,11pt]{article}

\newlength{\minitwocolumn}
\font\teneufm=eufm10
\font\seveneufm=eufm7
\font\fiveeufm=eufm5
\newfam\eufmfam
\textfont\eufmfam=\teneufm
\scriptfont\eufmfam=\seveneufm
\scriptscriptfont\eufmfam=\fiveeufm

\makeatletter
\@addtoreset{equation}{section}
\makeatother

\title{\bf
\Large{\bf
Free field approach to diagonalization of \\
boundary transfer matrix : recent advances}
}
\begin{document}
\maketitle
\begin{center}
{TAKEO KOJIMA}
\\~\\
{\it
Department of Mathematics and Physics,
Graduate School of Science and Engineering,\\
Yamagata University, Jonan 4-3-16, Yonezawa 992-8510, Japan}\\
kojima@yz.yamagata-u.ac.jp
\end{center}

~\\

\begin{abstract}
We diagonalize infinitely many 
commuting operators $T_B(z)$.
We call these operators $T_B(z)$
the boundary transfer matrix
associated with the quantum group and
the elliptic quantum group.
The boundary transfer matrix is related to 
the solvable model with a boundary.
When we diagonalize the boundary transfer matrix,
we can calculate the correlation functions
for the solvable model with a boundary.
We review 
the free field approach to
diagonalization of the boundary transfer matrix $T_B(z)$
associated with $U_q(A_2^{(2)})$ and
$U_{q,p}(\widehat{sl_N})$.
We construct the free field realizations of the eigenvectors
of the boundary transfer matrix $T_B(z)$.
This paper includes new unpublished formula of the eigenvector for
$U_q(A_2^{(2)})$.
It is thought that this diagonalization method
can be extended to more general quantum group
$U_q(g)$ and elliptic quantum group $U_{q,p}(g)$.
\end{abstract}

~\\

\newpage

\section{Introduction}

We study infinitely many 
commuting operators $T_B(z)$
that we call the boundary transfer matrix.
The boundary transfer matrix is related to 
the solvable model with a boundary.
There have been many developments in solvable models
in the last 30 years.
Various models were found to be exactly solvable and
various methods were invented to solve these models.
The Free field approach is a powerful method to study 
exactly solvable models \cite{JM}.
This paper is devoted to
the free field approach to diagonalization of
the boundary transfer matrix $T_B(z)$.
When we diagonalize the boundary transfer matrix,
we can calculate the correlation functions
for the solvable model with a boundary.
The first paper on this subject was
devoted to the XXZ chain with a boundary \cite{JKKKM},
in which the boundary transfer matrix $T_B(z)$
acts on the highest representation of 
the quantum group $U_q(\widehat{sl_2})$.
It is thought that this basic theory for the quantum group
$U_q(\widehat{sl_2})$ can be extended to the quantum group
$U_q(g)$ for arbitrary affine Lie algebra $g$.
It is thought that the theory on the quantum group $U_q(g)$
can be generalized to those on the elliptic quantum group $U_{q,p}(g)$.
In this paper we summarize the generalization on this direction. 
This paper includes a review on free field approach
to the boundary transfer matrix 
\cite{JKKKM, FK, YZ, Kojima2, MW, Kojima3, Kojima1}
and new unpublished formula of the boundary state
for the quantum group $U_q(A_2^{(2)})$.
The plan of this paper is as follows.
In section 2
we summarize the results for the quantum group
$U_q(A_2^{(2)})$ \cite{YZ} and give new unpublished formulae of
the boundary state associated with nontrivial K-matrix $K_\pm(z)$.
In section 3 we review the results for the elliptic quantum group 
$U_{q,p}(\widehat{sl_N})$, which gives a generalization of
the papers \cite{JKKKM, FK, MW, Kojima3}.

\section{Quantum group $U_q(A_2^{(2)})$}

In this section we diagonalize the boundary transfer matrix $T_B(z)$
for quantum group $U_q(A_2^{(2)})$.

\subsection{Boundary transfer matrix $T_B(z)$}

We fix $q$ and $z$ such that
$0<|q|<1$ and $|q^2|<|z|<|q^{-2}|$.
Let us set the q-integers 
$$[a]_q=\frac{q^a-q^{-a}}{q-q^{-1}}.$$
We use the abbreviation.
\begin{eqnarray}
(z;p_1,p_2,\cdots,p_M)_\infty=
\prod_{k_1,k_2,\cdots,k_M=0}^\infty
(1-p_1^{k_1}p_2^{k_2}\cdots p_M^{k_M}z).\nonumber
\end{eqnarray}
The R-matrix $R(\zeta)$ for the twisted quantum group
$U_q(A_2^{(2)})$ is given by following \cite{IK}.
\begin{eqnarray}
R(z)=\frac{1}{\kappa(z)}
\left(\begin{array}{ccccccccc}
1& & & & & & & & \\
 &b(z)& &c(z)& & & & & \\
 & &d(z)& & e(z)& &f(z)& &\\
 &zc(z)& &b(z)& & & & & \\
 & &-q^2z e(z)& &j(z)& &e(z)& \\
 & & & & &b(z)& &c(z)& \\
 & &n(z)& &-q^2z e(z)& &d(z)& &\\
 & & & & &zc(z)& &b(z)& \\
 & & & & & & & &1
\end{array}
\right).
\end{eqnarray}
Here we have set
\begin{eqnarray}
&&b(z)=\frac{q(z-1)}{q^2z-1},~c(z)=\frac{q^2-1}{q^2z-1},~
d(z)=\frac{q^2(z-1)(qz+1)}{
(q^2z-1)(q^3z+1)},~e(z)=
\frac{q^{\frac{1}{2}}(z-1)(q^2-1)}{(q^2z-1)(q^3z+1)},
\nonumber\\
&&f(z)=\frac{(q^2-1)((q^3+q)z-(q-1))}{
(q^2z-1)(q^3z+1)},~
n(z)=\frac{(q^2-1)((q^3-q)z+(q^2+1))}{
(q^2z-1)(q^3z+1)},\nonumber\\
&&j(z)=\frac{q^4z^2+(q^5-q^4-q^3+q^2+q-1)z-q}{
(q^2z-1)(q^3z+1)},\nonumber\\
&&\kappa(z)=z\frac{
(q^6z;q^6)_\infty
(q^2/z;q^6)_\infty
(-q^5z;q^6)_\infty
(-q^3/z;q^6)_\infty}{
(q^6/z;q^6)_\infty
(q^2z;q^6)_\infty
(-q^5/z;q^6)_\infty
(-q^3z;q^6)_\infty}.\nonumber
\end{eqnarray}
Let $\{v_+,v_0,v_-\}$
denote the natural basis of $V={\bf C}^3$.
When viewed as an operator on $V \otimes V$,
the matrix element of $R(z)$ are defined by
$R(z)v_{k_1}\otimes v_{k_2}=
\sum_{j_1,j_2=\pm,0}
v_{j_1}\otimes v_{j_2} R(z)_{j_1,j_2}^{k_1,k_2}$,
where the ordering of the index is given by
$(+,+),(+,0),(+,-),(0,+),(0,0),(0,-),(-,+),(-,0),(-,-)$.
The R-matrix $R(z)$ 
satisfies the Yang-Baxter equation.
\begin{eqnarray}
R_{1,2}(z_1/z_2)R_{1,3}(z_1/z_3)R_{2,3}(z_2/z_3)=
R_{2,3}(z_2/z_3)R_{1,3}(z_1/z_3)R_{1,2}(z_1/z_2).
\label{eqn:YBE1}
\end{eqnarray}
The R-matrix $R(z)$ is characterized by
the intertwiner as the quantum group $U_q(A_2^{(2)})$.
We set the normalization function $\kappa(z)$
such that the minimal eigenvalue of the corner 
transfer matrix becomes $1$ \cite{Baxter}.
There exist three diagonal solutions 
of the K-matrix for the
quantum group $U_{q}(A_2^{(2)})$.
The K-matrix $K_\epsilon(z)$,
$(\epsilon=\pm,0)$ are given by following \cite{BFKZ}.
\begin{eqnarray}
K_0(z)=\frac{\varphi_0(z)}{\varphi_0(z^{-1})}
\left(\begin{array}{ccc}
1& & \\
 &1& \\
 & &1
\end{array}\right),~~~
K_\pm (z)=
\frac{\varphi_\pm (z)}{\varphi_\pm (z^{-1})}
\left(\begin{array}{ccc}
z^2& & \\
 &\frac{\displaystyle
\pm \sqrt{-1}q^{\frac{3}{2}}+z}{
\displaystyle
\pm \sqrt{-1}q^{\frac{3}{2}}+z^{-1}}& \\
 & &1
\end{array}\right).
\label{def:K-matrix2}
\end{eqnarray}
where 
\begin{eqnarray}
\varphi_0(z)=
\frac{
(q^8z;q^{12})_\infty 
(-q^9z^2;q^{12})_\infty}{
(q^{12}z;q^{12})_\infty 
(-q^5z^2;q^{12})_\infty},
\varphi_\pm(z)=
\frac{
(\pm \sqrt{-1}q^{\frac{9}{2}}z;q^6)_\infty 
(\mp \sqrt{-1}q^{\frac{7}{2}}z;q^6)_\infty}{
(\pm \sqrt{-1}q^{\frac{1}{2}}z;q^6)_\infty 
(\mp \sqrt{-1}q^{\frac{3}{2}}z;q^6)_\infty}
~\varphi_0(z).
\end{eqnarray}
The K-matrix $K(z)=K_\epsilon(z)$ satisfies the boundary
Yang-Baxter equation in ${\rm End}(V \otimes V)$ \cite{Sklyanin}.
\begin{eqnarray}
K_2(z_2)R_{2,1}(z_1 z_2)
K_1(z_1)R_{1,2}(z_1/z_2)=
R_{2,1}(z_1/z_2)K_1(z_1)
R_{1,2}(z_1 z_2)K_2(z_2).
\label{eqn:boundaryYBE1}
\end{eqnarray}
We set the normalization function 
$\varphi_\epsilon(z)$, $(\epsilon=\pm,0)$ such that
the minimal eigenvalue of the boundary transfer matrix 
$T_B(z)$ becomes 1.
The Izergin-Korepin model
associated with the identity solution 
$\bar{K}_0(z)=id$ was studied in \cite{YZ}.
In this paper we give the free field realization
of the boundary state
for nontrivial solutions $K_\pm(z)$.
The Izergin-Korepin model with
the nontrivial solutions $K_\epsilon(z)$, 
$(\epsilon=\pm,0)$
was studied in \cite{Kojima2}.

Let us introduce the vertex operators 
$\Phi_\epsilon(z)$, $(\epsilon=\pm,0)$ \cite{Matsuno, JingMisra, HYZ}
that satisfy
the following commutation relation
\begin{eqnarray}
\Phi_{\epsilon_2}(z_2)\Phi_{\epsilon_1}(z_1)=
\sum_{\epsilon_1', \epsilon_2'=\pm, 0}
R_{\epsilon_1 \epsilon_2}^{\epsilon_1' \epsilon_2'}
(z_1/z_2)
\Phi_{\epsilon_1'}(z_1)\Phi_{\epsilon_2'}(z_2).
\label{eqn:VO}
\end{eqnarray}
The vertex operator $\Phi_\epsilon(z)$ and its dual
$\Phi_\epsilon^*(z)=q^{-\frac{\epsilon}{2}}
\Phi_{-\epsilon}(-q^{-3}z)$, $(\epsilon=\pm,0)$
satisfy the inversion relation
\begin{eqnarray}
g\Phi_{\epsilon_1}(z)\Phi_{\epsilon_2}^*(z)=
\delta_{\epsilon_1, \epsilon_2} id,
\label{eqn:VO2}
\end{eqnarray}
where 
$g=\frac{1}{1+q}\frac{(q^2;q^6)_\infty
(-q^3;q^6)_\infty}{
(q^6;q^6)_\infty
(-q^5;q^6)_\infty}$. 
Let us introduce the boundary transfer matrix $T_B(z)$ 
for $U_q(A_2^{(2)})$.
The boundary transfer matrix $T_B(z)$ is given by
\begin{eqnarray}
T_B(z)=g \sum_{\epsilon,\epsilon'=\pm, 0}
\Phi_\epsilon^*(z^{-1})K_\epsilon^{\epsilon'}(z)
\Phi_{\epsilon'}(z).
\end{eqnarray}
The boundary transfer matrix $T_B(z)$
is related to the boundary Izergin-Korepin model
\cite{YZ, Kojima2}.
From the commutation relations of the vertex operators
(\ref{eqn:VO}) and the boundary Yang-Baxter equation
(\ref{eqn:boundaryYBE1}),
the commutativity of the boundary transfer matrix
$T_B(z)$ is ensured.
\begin{eqnarray}
~[T_B(z_1),T_B(z_2)]=0,~~
{\rm for~any~~} z_1,z_2.
\end{eqnarray}
We call the eigenvector $|B\rangle_\epsilon$,$(\epsilon=\pm,0)$ 
with eigenvalue $1$
the boundary state.
\begin{eqnarray}
T_B(z)|B\rangle_\epsilon=|B\rangle_\epsilon.
\end{eqnarray}
From the definition of $\varphi_\epsilon(z)$, 
the boundary state
$|B\rangle$
is the eigenvector with the minimal
eigenvalue.
In the following section we give the free field realization
of the boundary state $|B\rangle_\epsilon$ for
the diagonal boundary K-matrix $K_\epsilon(z)$, 
$(\epsilon=\pm, 0)$.
The boundary state $|B \rangle_\epsilon$
for the identity K-matrix $K_0(z)$ was constructed in \cite{YZ}.
Let us introduce the type-II vertex operators 
$\Psi_\mu^*(\xi)$, $(\mu=\pm,0)$ 
that satisfy the following commutation relation.
\begin{eqnarray}
\Phi_\epsilon(z)\Psi_\mu^*(\xi)=\tau(z/\xi)
\Psi_\mu^*(\xi)\Phi_\epsilon(z),~~~(\epsilon,\mu=\pm,0),
\end{eqnarray}
where 
\begin{eqnarray}
\tau(z)=z^{-1}\frac{
\Theta_{q^6}(q^5z)
\Theta_{q^6}(-q^4z)}
{\Theta_{q^6}(q^5z^{-1})
\Theta_{q^6}(-q^4z^{-1})},~~
\Theta_p(z)=(p;p)_\infty (z;p)_\infty (pz^{-1};p)_\infty.\nonumber
\end{eqnarray}
Multiplying the type-II vertex operators
$\Psi_\mu^*(\xi)$ to the boundary state $|B\rangle_\epsilon$,
\begin{eqnarray}
|\xi_1,\xi_2,\cdots,\xi_N
\rangle_{\mu_1,\mu_2,\cdots, \mu_N}=
\Psi_{\mu_1}^*(\xi_1)\Psi_{\mu_2}^*(\xi_2)
\cdots \Psi_{\mu_N}^*(\xi_N)|B\rangle_\epsilon,
\end{eqnarray}
we have many eigenvectors.
\begin{eqnarray}
T_B(z)|\xi_1,\xi_2,\cdots,\xi_N
\rangle_{\mu_1,\mu_2,\cdots,\mu_N}
=\prod_{s=1}^N 
\tau(z/\xi_s)\tau(-1/q^3z\xi_s)
|\xi_1,\xi_2,\cdots,\xi_N\rangle_{\mu_1,\mu_2,\cdots,\mu_N}.\nonumber
\end{eqnarray}
The vectors 
$\{|\xi_1,\cdots,\xi_N\rangle_{\mu_1,\cdots,\mu_N}\}$
are the basis of the space of the state
of the boundary Izergin-Korepin model.

\subsection{Free field realization}
In this section we give the free field realization
of the boundary state $|B\rangle_\epsilon$.
Let us introduce bosons $a_m$,$(m\in {\bf Z}_{\neq 0})$ as following
\cite{Matsuno, Jing, JingMisra, HYZ}.
\begin{eqnarray}
~[a_m,a_n]=\delta_{m+n}\frac{[m]_q}{m}([2m]_q-(-1)^m[m]_q).
\end{eqnarray}
Let us set the zero-mode operators $P,Q$ by
\begin{eqnarray}
~[a_m,P]=[a_m,Q]=0,~[P,Q]=1.
\end{eqnarray}
Level-$1$ irreducible highest representation $V(\Lambda_1)$
of $U_q(A_2^{(2)})$ is realized by
\begin{eqnarray}
V(\Lambda_1)={\bf C}[a_{-1},a_{-2},\cdots]
\oplus_{n \in {\bf Z}}e^{nQ}|\Lambda_1 \rangle,~~
|\Lambda_1\rangle=e^{\frac{Q}{2}}|0\rangle.\nonumber
\end{eqnarray}
The vacuum vector $|0\rangle$ is characterized by
\begin{eqnarray}
a_m|0\rangle=0,~(m>0),~~P|0\rangle=0. \nonumber
\end{eqnarray}
Let us set the auxiliary operators 
$P(z),Q(z),R^-(w),S^-(w)$ by
\begin{eqnarray}
P(z)=\sum_{m>0}\frac{a_{-m}
q^{\frac{9m}{2}}z^m}{[2m]_q-(-1)^m[m]_q},&&
Q(z)=-\sum_{m>0}\frac{a_m
q^{-\frac{7m}{2}}z^{-m}
}{[2m]_q-(-1)^m[m]_q},
\nonumber
\\
R^-(w)=-\sum_{m>0}\frac{a_{-m}}{[m]_q}
q^{\frac{m}{2}}w^m,
&&
S^-(w)=\sum_{m>0}\frac{a_m}{[m]_q}
q^{\frac{m}{2}}w^{-m}.\nonumber
\end{eqnarray}
Let us set 
$\epsilon(q)=
([2]_{q^{\frac{1}{2}}})^{\frac{1}{2}}$.
Let us set the current,
\begin{eqnarray}
X^-(w)=\epsilon(q)e^{R^-(w)}e^{S^-(w)}
e^{-Q}w^{-P+\frac{1}{2}}.\nonumber
\end{eqnarray}
The free field realizations of the vertex operators 
$\Phi_\epsilon(z)$ \cite{Matsuno, JingMisra, HYZ} are given by
\begin{eqnarray}
\Phi_-(z)&=&\frac{1}{\epsilon(q)}
e^{P(z)}e^{Q(z)}e^Q(-zq^4)^{P+\frac{1}{2}},\nonumber\\
\Phi_0(z)&=&\oint_{C_1} \frac{dw}{2\pi \sqrt{-1}w}
\frac{(q^2-1)}{q^4z (1-qw/z)(1-qz/w)}:\Phi_-(z)X^-(q^4w):,\nonumber\\
\Phi_+(z)&=&\oint \oint_{C_2} 
\frac{dw_1}{2\pi \sqrt{-1}w_1}
\frac{dw_2}{2\pi \sqrt{-1}w_2}
\frac{q^{1/2}(1-q^2)^2}{q^4z^2w_1w_2}\nonumber\\
&\times&
\frac{(w_1-w_2)^2 (q(w_1+w_2)-(1+q^2)z)}{
(1+qw_1/w_2)(1+qw_2/w_1)
(1-qw_1/z)(1-qz/w_1)(1-qw_2/z)(1-qz/w_2)}\nonumber\\
&\times&
:\Phi_-(z)X^-(q^4w_1)X^-(q^4w_2):.\nonumber
\end{eqnarray}
The integrand contour $C_1$
encircles $w=0,qz$ but not $w=q^{-1}z$.
The integrand contour $C_2$
encircles $w_1=0,qz,qw_2$ and $w_2=0,qz,qw_1$ 
but not $w_1=q^{-1}z,q^{-1}w_2$ and $w_2=q^{-1}z,q^{-1}w_1$.
The free field realization of type-II vertex operators
$\Psi_\mu^*(\xi)$ are given as similar way \cite{Matsuno}.
Now we have the free field realization
of the boundary transfer matrix $T_B(z)$,
using those of the vertex operators.
We construct
the free field realization of 
the boundary state $|B\rangle_\epsilon$,
analyzing those of the
boundary transfer matrix $T_B(z)$.
The following is {\bf main result} of this section.
The free field realization 
of the boundary states $|B\rangle_\epsilon$,
$(\epsilon=\pm,0)$ are given by
\begin{eqnarray}
|B\rangle_\epsilon=e^{F_\epsilon}
e^{-\frac{Q}{2}}|0\rangle,~~
(\epsilon=\pm,0).
\end{eqnarray}
Here we have set
\begin{eqnarray}
F_\epsilon
&=&-\frac{1}{2}\sum_{m>0}\frac{mq^{8m}}{
[2m]_q-(-1)^m[m]_q}a_{-m}^2\nonumber\\
&+&\sum_{m>0}
\left\{\theta_m\left(\frac{
(q^{\frac{m}{2}}-q^{-\frac{m}{2}}-\sqrt{-1}^{~m})q^{4m}}{[2m]_q-(-1)^m[m]_q}\right)
-\frac{(\epsilon \sqrt{-1})^{m} q^{3m}}
{[2m]_q-(-1)^m[m]_q}\right\} a_{-m}.
\end{eqnarray}
Here we have used 
$\theta_m(x)=\left\{\begin{array}{cc}
x,&m:{\rm even}\\
0,&m:{\rm odd}
\end{array}\right.$.
The boundary state $|B\rangle_0$ for the identity K-matrix
$\bar{K}_0(z)=id$ was constructed in \cite{YZ}.
The realizations of $|B\rangle_\epsilon$ 
for the nontrivial K-matrix $K_\epsilon(z)$ are new.
Multiplying the type-II vertex operators $\Psi_\mu^*(\xi)$
to the boundary state $|B\rangle_\epsilon$,
we get the diagonalization of
the boundary transfer matrix $T_B(z)$.
It is thought that 
this method can be extended to the case of 
the affine quantum group $U_q(g)$.

\section{Elliptic quantum group 
$U_{q,p}(\widehat{sl_N})$}

In this section
we diagonalize the boundary transfer matrix $T_B(z)$
associated with the elliptic quantum group 
$U_{q,p}(\widehat{sl_N})$ \cite{Kojima1}. 
It gives a generalization of the papers
\cite{JKKKM, FK, MW, Kojima3}.

\subsection{Boundary transfer matrix}

Let us set the integer $N=2,3,\cdots$.
We assume that $0<x<1$ and $r \geq N+2~(r \in {\bf Z})$.
We set
$z=x^{2u}, x=e^{-\pi i/r \tau}$.
We set the elliptic theta function 
$[u]$ by
\begin{eqnarray}
~[u]=x^{\frac{u^2}{r}-u}\Theta_{x^{2r}}(x^{2u}),~~~
\Theta_q(z)&=&
(q;q)_\infty (z;q)_\infty (q/z;q)_\infty.\nonumber
\end{eqnarray}
Let $\epsilon_\mu (1\leq \mu \leq N)$ be the orthonormal
basis of ${\bf R}^N$ with the inner
product $(\epsilon_\mu |\epsilon_\nu)=\delta_{\mu,\nu}$.
Let us set $\bar{\epsilon}_\mu=\epsilon_\mu-\epsilon$
where $\epsilon=\frac{1}{N}\sum_{\nu=1}^N \epsilon_\nu$.
Note that $\sum_{\mu=1}^N \bar{\epsilon}_\mu=0$.
Let $\alpha_\mu~(1\leq \mu \leq N-1)$ the simple root :
$\alpha_\mu=\bar{\epsilon}_\mu-\bar{\epsilon}_{\mu+1}$.
Let $\omega_\mu~(1\leq \mu \leq N-1)$ be 
the fundamental weights, which satisfy
\begin{eqnarray}
(\alpha_\mu|\omega_\nu)=\delta_{\mu,\nu},
~~(1\leq \mu,\nu\leq N-1).\nonumber
\end{eqnarray}
Explicitly we set
$\omega_\mu=\sum_{\nu=1}^\mu \bar{\epsilon}_\nu.$
The type $A_{N-1}$ weight lattice is the
linear span of $\bar{\epsilon}_\mu$ or $\omega_\mu$.
\begin{eqnarray}
P=\sum_{\mu=1}^{N-1} {\bf Z}\bar{\epsilon}_\mu
=\sum_{\mu=1}^{N-1} {\bf Z}\omega_\mu.\nonumber
\end{eqnarray}
For $a \in P$ we set $a_\mu$ and $a_{\mu,\nu}$ by
\begin{eqnarray}
a_{\mu,\nu}=a_{\mu}-a_{\nu},~~~
a_{\mu}=(a+\rho|\bar{\epsilon}_\mu),~~~(\mu, \nu \in P).
\nonumber
\end{eqnarray}
Here we set
$\rho=\sum_{\mu=1}^{N-1}\omega_\mu$.
Let us set the restricted path $P_{r-N}^+$ by
\begin{eqnarray}
P_{r-N}^+=\{a=\sum_{\mu=1}^{N-1}c_\mu \omega_\mu \in P|
c_\mu \in {\bf Z}, c_\mu \geq 0, 
\sum_{\mu=1}^{N-1}c_\mu \leq r-N \}.
\nonumber
\end{eqnarray}
For $a \in P_{r-N}^+$, condition 
$0<a_{\mu,\nu}<r,~(1\leq \mu<\nu\leq N-1)$ holds.

We recall elliptic solutions of
the Yang-Baxter equation of face type.
An ordered pair $(b,a)\in P^2$ is called
admissible if and only if there exists
$\mu~(1\leq \mu \leq N)$ such that
$b-a=\bar{\epsilon}_\mu$.
An ordered set of four weights $(a,b,c,d)\in P^4$
is called an admissible configuration
around a face if and only if
the ordered pairs $(b,a)$, $(c,b)$, $(d,a)$ and $(c,d)$
are admissible.
Let us set
the Boltzmann weight functions 
$W\left(\left.\begin{array}{cc}
c&d\\
b&a
\end{array}\right|u\right)$
associated with admissible configuration
$(a,b,c,d)\in P^4$ \cite{JMO}.
For $a \in P_{r-N}^+$, we set
\begin{eqnarray}
&&W\left(\left.
\begin{array}{cc}
a+2\bar{\epsilon}_\mu & a+\bar{\epsilon}_\mu\\
a+\bar{\epsilon}_\mu & a
\end{array}\right|u\right)=R(u),\\
&&W\left(\left.
\begin{array}{cc}
a+\bar{\epsilon}_\mu+\bar{\epsilon}_\nu & 
a+\bar{\epsilon}_\mu\\
a+\bar{\epsilon}_\nu & a
\end{array}\right|u\right)=R(u)\frac{[u][a_{\mu,\nu}-1]}
{[u-1][a_{\mu,\nu}]},\\
&&W\left(\left.
\begin{array}{cc}
a+\bar{\epsilon}_\mu+\bar{\epsilon}_\nu 
& a+\bar{\epsilon}_\nu\\
a+\bar{\epsilon}_\nu & a
\end{array}\right|u\right)=
R(u)\frac{[u-a_{\mu,\nu}][1]}{
[u-1][a_{\mu,\nu}]}.
\end{eqnarray}
The normalizing function $R(u)$ is 
given by 
\begin{eqnarray}
R(u)&=&z^{\frac{r-1}{r}\frac{N-1}{N}}
\frac{\varphi(z^{-1})}{\varphi(z)},~~~
\varphi(z)=\frac{
(x^{2}z;x^{2r},x^{2N})_\infty 
(x^{2r+2N-2}z;x^{2r},x^{2N})_\infty}{
(x^{2r}z;x^{2r},x^{2N})_\infty 
(x^{2N}z;x^{2r},x^{2N})_\infty}.
\nonumber
\end{eqnarray}
Because $0<a_{\mu,\nu}<r~(1\leq \mu<\nu \leq N-1)$
holds for $a \in P_{r-N}^+$,
the Boltzmann weight functions
are well defined.
The Boltzmann weight functions 
satisfy the Yang-Baxter equation of the face type.
\begin{eqnarray}
&&\sum_{g}
W\left(\left.\begin{array}{cc}
d&e\\
c&g
\end{array}
\right|u_1\right)
W\left(\left.\begin{array}{cc}
c&g\\
b&a
\end{array}
\right|u_2\right)
W\left(\left.\begin{array}{cc}
e&f\\
g&a
\end{array}
\right|u_1-u_2\right)
\nonumber\\
&=&
\sum_{g}
W\left(\left.\begin{array}{cc}
g&f\\
b&a
\end{array}
\right|u_1\right)
W\left(\left.\begin{array}{cc}
d&e\\
g&f
\end{array}
\right|u_2\right)
W\left(\left.\begin{array}{cc}
d&g\\
c&b
\end{array}
\right|u_1-u_2\right).
\label{eqn:Boltzmann1}
\end{eqnarray}
We set the normalization function 
$\varphi(z)$ such that the minimal eigenvalue of 
the corner transfer matrix becomes 1
\cite{Baxter}.
An order set of three weights $(a,b,g)
\in P^3$
is called 
an admissible configuration at a boundary
if and only if
the ordered pairs $(g,a)$ and $(g,b)$ are admissible. 
Let us set the boundary Boltzmann weight functions
$
K\left(
\left.\begin{array}{cc}
&a\\
g&\\
&b
\end{array}
\right|u\right)
$ for admissible weights $(a,b,g)$ as following \cite{BFKZ}.
\begin{eqnarray}
K\left(
\left.\begin{array}{cc}
&a\\
a+\bar{\epsilon}_\mu&\\
&b
\end{array}
\right|u\right)=
z^{\frac{r-1}{r}\frac{N-1}{N}-\frac{2}{r}
a_1}\frac{h(z)}{h(z^{-1})}
\frac{[c-u][a_{1,\mu}+c+u]}
{[c+u][a_{1,\mu}+c-u]}
\delta_{a,b}.
\end{eqnarray}
In this paper, we consider
the case of continuous parameter $0<c<1$.
The normalization function
$h(z)$ is given by following \cite{Kojima1}.
\begin{eqnarray}
h(z)&=&
\frac{
(x^{2r+2N-2}/z^2;x^{2r},x^{4N})_\infty 
(x^{2N+2}/z^2;x^{2r},x^{4N})_\infty}{
(x^{2r}/z^2;x^{2r},x^{4N})_\infty 
(x^{4N}/z^2;x^{2r},x^{4N})_\infty}\nonumber\\
&\times&
\frac{
(x^{2N+2c}/z;x^{2r},x^{2N})_\infty
(x^{2r-2c}/z;x^{2r},x^{2N})_\infty}{
(x^{2N+2r-2c-2}/z;x^{2r},x^{2N})_\infty
(x^{2c+2}/z;x^{2r},x^{2N})_\infty}\nonumber\\
&\times&
\prod_{j=2}^N 
\frac{
(x^{2r+2N-2c-2a_{1,j}}/z;x^{2r},x^{2N})_\infty
(x^{2c+2a_{1,j}}/z;x^{2r},x^{2N})_\infty}{
(x^{2r+2N-2c-2a_{1,j}-2}/z;x^{2r},x^{2N})_\infty
(x^{2c+2+2a_{1,j}}/z;x^{2r},x^{2N})_\infty}.
\end{eqnarray}
The boundary Boltzmann weight functions 
and the Boltzmann weight functions
satisfy the Boundary Yang-Baxter equation.
\begin{eqnarray}
&&\sum_{f,g}
W\left(\left.\begin{array}{cc}
c&f\\
b&a
\end{array}\right|u_1-u_2\right)
W\left(\left.\begin{array}{cc}
c&d\\
f&g
\end{array}
\right|u_1+u_2\right)
K\left(\left.\begin{array}{cc}
~&g\\
f&\\
~&a
\end{array}
\right|u_1\right)
K\left(\left.\begin{array}{cc}
~&e\\
d&\\
~&g
\end{array}
\right|u_2\right)
\nonumber\\
&=&
\sum_{f,g}
W\left(\left.\begin{array}{cc}
c&d\\
f&e
\end{array}\right|u_1-u_2\right)
W\left(\left.\begin{array}{cc}
c&f\\
b&g
\end{array}
\right|u_1+u_2\right)
K\left(\left.\begin{array}{cc}
~&e\\
f&\\
~&g
\end{array}
\right|u_1\right)
K\left(\left.\begin{array}{cc}
~&g\\
b&\\
~&a
\end{array}
\right|u_2\right).
\label{eqn:boundaryYBE2}
\end{eqnarray}
We set the normalization function $h(z)$ such that
the minimal eigenvalue of the boundary transfer matrix $T_B(z)$
becomes $1$.

The vertex operator $\Phi^{(b,a)}(z)$ 
and the dual vertex operator $\Phi^{*(a,b)}(z)$
associated with the elliptic quantum group
$U_{q,p}(\widehat{sl_N})$,
are the operators
which satisfy the following commutation relations
\begin{eqnarray}
\Phi^{(a,b)}(z_1)
\Phi^{(b,c)}(z_2)
&=&\sum_{g}
W\left(\left.\begin{array}{cc}
a&g\\
b&c
\end{array}
\right|u_2-u_1\right)
\Phi^{(a,g)}(z_2)
\Phi^{(g,c)}(z_1),\label{eqn:VO1}\\
\Phi^{*(a,b)}(z_1)
\Phi^{*(b,c)}(z_2)
&=&
\sum_{g}
W\left(\left.\begin{array}{cc}
c&b\\
g&a
\end{array}
\right|u_2-u_1\right)
\Phi^{*(a,g)}(z_2)
\Phi^{*(g,c)}(z_1),\label{eqn:VO2}\\
\Phi^{(a,b)}(z_1)
\Phi^{*(b,c)}(z_2)
&=&
\sum_{g}
W\left(\left.\begin{array}{cc}
g&c\\
a&b
\end{array}
\right|u_1-u_2\right)
\Phi^{*(a,g)}(z_2)
\Phi^{(g,c)}(z_1),\label{eqn:VO3}
\end{eqnarray}
and the inversion relation
\begin{eqnarray}
\Phi^{(a,g)}(z)\Phi^{*(g,b)}(z)=\delta_{a,b}.
\label{eqn:inversion1}
\end{eqnarray}
We define the boundary transfer matrix $T_B(z)$
for the elliptic quantum group $U_{q,p}(\widehat{sl_N})$.
\begin{eqnarray}
T_B(z)=\sum_{\mu=1}^N
\Phi^{*(a,a+\bar{\epsilon}_\mu)}(z^{-1})
K\left(\left.
\begin{array}{cc}
~& a\\
a+\bar{\epsilon}_\mu &\\
~& a
\end{array}
\right|u\right)
\Phi^{(a+\bar{\epsilon}_\mu,a)}(z).
\label{def:boundary-transfer}
\end{eqnarray}
From the commutation relations of the vertex operators (\ref{eqn:VO1}),
(\ref{eqn:VO2}), (\ref{eqn:VO3}), and the boundary Yang-Baxter equation
(\ref{eqn:boundaryYBE2}),
the boundary $T_B(z)$ commute with each other.
\begin{eqnarray}
~[T_B(z_1),T_B(z_2)]=0,~~~{\rm for~any~}z_1, z_2.
\end{eqnarray}
We call the eigenvector $|B\rangle$ with the eigenvalue $1$
the boundary state.
\begin{eqnarray}
T_B(z)|B\rangle=|B\rangle.
\end{eqnarray}
Let us introduce the type-II vertex operators
$\Psi^{*(b,a)}(z)$ by
\begin{eqnarray}
\Phi^{(d,c)}(z_1)\Psi^{*(b,a)}(z_2)=
\chi(z_2/z_1)
\Psi^{* (b,a)}(z_2)\Phi^{(d,c)}(z_1),
\label{eqn:IIVO2}\\
\Phi^{*(c,d)}(z_1)\Psi^{*(b,a)}(z_2)=
\chi(z_1/z_2)
\Psi^{* (b,a)}(z_2)\Phi^{*(c,d)}(z_1).
\label{eqn:IIVO3}
\end{eqnarray}
where we have set
\begin{eqnarray}
\chi(z)=z^{-\frac{N-1}{N}}
\frac{\Theta_{x^{2N}}(-xz)}{
\Theta_{x^{2N}}(-xz^{-1})}.
\nonumber
\end{eqnarray}
We set the vectors $|\xi_1,\xi_2,\cdots,\xi_M
\rangle_{\mu_1,\mu_2,\cdots,\mu_M}$
$(1\leq \mu_1,\mu_2,\cdots,\mu_M \leq N)$. 
\begin{eqnarray}
&&
|\xi_1,\xi_2,\cdots,\xi_M
\rangle_{\mu_1,\mu_2,\cdots,\mu_M}
\label{def:excitations}
\\
&=&
\Psi^{*(b+\bar{\epsilon}_{\mu_1}
+\bar{\epsilon}_{\mu_2}+\cdots
+\bar{\epsilon}_{\mu_M},
b+\bar{\epsilon}_{\mu_2}+\cdots
+\bar{\epsilon}_{\mu_M})}(\xi_1)
\cdots
\Psi^{*(b+\bar{\epsilon}_{\mu_{M-1}}
+\bar{\epsilon}_{\mu_M},b+\bar{\epsilon}_{\mu_M})}(\xi_{M-1})
\Psi^{*(b+\bar{\epsilon}_{\mu_M},b)}(\xi_M)
|B \rangle.
\nonumber
\end{eqnarray}
Now we have many eigenvectors of $T_B(z)$.
\begin{eqnarray}
T_B(z)|\xi_1,\xi_2,\cdots,\xi_M
\rangle_{\mu_1,\mu_2,\cdots,\mu_M}
=
\prod_{j=1}^M \chi(\xi_j/z)\chi(1/\xi_j z)~
|\xi_1,\xi_2,\cdots,\xi_M
\rangle_{\mu_1,\mu_2,\cdots,\mu_M}.\nonumber
\end{eqnarray}
The vectors $
|\xi_1,\xi_2,\cdots,\xi_M
\rangle_{\mu_1,\mu_2,\cdots,\mu_M}$
are the basis of the space of the state of 
the boundary $U_{q,p}(\widehat{sl_N})$ face model.

\subsection{Free field realization}

In this section we give the free field realizations
of the boundary state $|B\rangle$.
Let us introduce the bosons 
$\beta_m^i, (i=1,2,\cdots,N-1;m \in {\bf Z})$ as following \cite{AJMP}.
\begin{eqnarray}
~[\beta_m^j,\beta_n^k]=
\left\{
\begin{array}{cc}
\displaystyle
m \frac{[(r-1)m]_x }{[rm]_x}
\frac{[(N-1)m]_x}{[Nm]_x}\delta_{m+n,0}
&(j=k)\\
\displaystyle
-m x^{Nm~sgn(j-k)}
\frac{[(r-1)m]_x}{[rm]_x}
\frac{[m]_x}{[Nm]_x}\delta_{m+n,0}
&(j \neq k).
\end{array}
\right.
\end{eqnarray}
Let us set $\beta_m^N$ by
$\sum_{j=1}^Nx^{-2jm}\beta_m^j=0$.
The above commutation relations
are valid for all $1\leq j,k \leq N$.
We also introduce
the zero-mode operators 
$P_\alpha, Q_\alpha$, $(\alpha \in P)$ by
\begin{eqnarray}
~[\sqrt{-1}P_\alpha,Q_\beta]=(\alpha|\beta),~~(\alpha,\beta \in P).
\end{eqnarray}
In what follows we 
deal with the bosonic Fock space 
${\cal F}_{l,k}$, generated by 
$\beta_{-m}^j (m>0)$ over the vacuum vector 
$|l,k \rangle$, where
$l=b+\rho,~k=a+\rho$
for $a \in P_{r-N}^+, 
b \in P_{r-1-N}^+$.
\begin{eqnarray}
{\cal F}_{l,k}={\bf C}[\{\beta_{-1}^j,
\beta_{-2}^j,\cdots\}_{j=1,\cdots,N-1}]|l,k\rangle,
~~|l,k\rangle=e^{\sqrt{-1}\sqrt{\frac{r}{r-1}}Q_l-
\sqrt{-1}\sqrt{\frac{r-1}{r}}Q_k}|0,0\rangle.\nonumber
\end{eqnarray}
where
\begin{eqnarray}
&&\beta_m^j |l,k\rangle=0,~(m>0),~
P_\alpha |l,k\rangle=\left(\alpha\left|
\sqrt{\frac{r}{r-1}}l-\sqrt{\frac{r-1}{r}}k
\right.\right)|l,k\rangle.\nonumber
\end{eqnarray}
The commutation relation of bosons $\beta_m^j$
is not symmetric.
It is convenient to introduce new generators
of bosons $\alpha_m^j~(m \in {\bf Z}_{\neq 0};
1\leq j \leq N-1)$ by
\begin{eqnarray}
\alpha_m^j=x^{-jm}(\beta_m^j-\beta_m^{j+1}).
\end{eqnarray}
They satisfy the following commutation relations.
\begin{eqnarray}
~[\alpha_m^j,\alpha_n^k]=m\frac{[(r-1)m]_x}{[rm]_x}
\frac{[A_{j,k}m]_x}{[m]_x}\delta_{m+n,0},\nonumber
\end{eqnarray}
where $A_{j,k}$ is a matrix element of the Cartan matrix of
${sl_N}$ type.
We give a free field realization of
the vertex operators $\Phi^{(b,a)}(z)$.
Let us set the operators 
$P_-(z),Q_-(z)$,
$R_-^j(z),S_-^j(z)$, 
$(1\leq j \leq N-1)$ 
by
\begin{eqnarray}
P_-(z)&=&\sum_{m>0}\frac{1}{m}\beta_{-m}^1z^{m},~
Q_-(z)=-\sum_{m>0}\frac{1}{m}\beta_m^1z^{-m},
\nonumber
\\
R_-^j(z)&=&
-\sum_{m>0}\frac{1}{m}\alpha_{-m}^j z^m,
~S_-^j(z)=
\sum_{m>0}\frac{1}{m}\alpha_m z^{-m}.\nonumber
\end{eqnarray}
Let us set the basic operators $U(z)$,$F_{\alpha_j}(z)$,
$(1\leq j \leq N-1)$
on the Fock space ${\cal F}_{l,k}$.
\begin{eqnarray}
U(z)&=&
z^{\frac{r-1}{2r}\frac{N-1}{N}}
e^{-\sqrt{-1}\sqrt{\frac{r-1}{r}}Q_{\bar{\epsilon}_1}}
z^{-\sqrt{\frac{r-1}{r}}P_{\bar{\epsilon}_1}}
e^{P_-(z)}e^{Q_-(z)},\nonumber
\\
F_{\alpha_j}(z)&=&
z^{\frac{r-1}{r}}
e^{\sqrt{-1}\sqrt{\frac{r-1}{r}}Q_{\alpha_j}}
z^{\sqrt{\frac{r-1}{r}}P_{\alpha_j}}
e^{R_-^j(z)}e^{S_-^j(z)}.\nonumber
\end{eqnarray}
In what follows we set $l=b+\rho, k=a+\rho$, 
$(a \in P_{r-N}^+, b \in P_{r-N-1}^+)$
and $\pi_\mu=\sqrt{r(r-1)}P_{\bar{\epsilon}_\mu},~
\pi_{\mu, \nu}=\pi_\mu-\pi_\nu$.
Then $\pi_{\mu \nu}$ acts on ${\cal F}_{l,k}$
as an integer $(\epsilon_\mu-\epsilon_\nu|rl-(r-1)k)$.
We give the free field realization of
the vertex operators
$\Phi^{(a+\bar{\epsilon}_{\mu},a)}(z)$, 
$(1\leq \mu \leq N-1)$ \cite{AJMP} by
\begin{eqnarray}
\Phi^{(a+\bar{\epsilon}_1,a)}
(z_0^{-1})&=&
U(z_0),\nonumber
\\
\Phi^{(a+\bar{\epsilon}_\mu,a)}(z_0^{-1})
&=&
\oint
\cdots \oint 
\prod_{j=1}^{\mu-1}
\frac{dz_j}{2\pi i z_j} U(z_0)
F_{\alpha_1}(z_1)F_{\alpha_2}(z_2)
\cdots
F_{\alpha_{\mu-1}}(z_{\mu-1})\nonumber\\
&\times&
\prod_{j=1}^{\mu-1}
\frac{[u_j-u_{j-1}+\frac{1}{2}-\pi_{j,\mu}]}
{[u_j-u_{j-1}-\frac{1}{2}]}.
\nonumber
\end{eqnarray}
Here we set $z_j=x^{2u_j}$.
We take the integration contour to be simple closed 
curve that encircles 
$z_j=0, x^{1+2rs}z_{j-1}, (s \in {\bf N})$
but not
$z_j=x^{-1-2rs}z_{j-1}, (s \in {\bf N})$
for $1\leq j \leq \mu-1$.
The $\Phi^{(a+\bar{\epsilon}_\mu,a)}(z)$ 
is an operator such that
$\Phi^{(a+\bar{\epsilon}_\mu,a)}(z):
{\cal F}_{l,k}\to {\cal F}_{l,k+\bar{\epsilon}_\mu}$.
The free field realization of the dual vertex operator
$\Phi^{*(a,b)}(z)$ and the type-II vertex operator $\Psi^{*(a,b)}(z)$
are given by similar way \cite{AJMP, FKQ}.
Now we have the free field realization
of the boundary transfer matrix $T_B(z)$,
using those of the vertex operators.
We construct
the free field realization of 
the boundary state $|B\rangle$,
analyzing those of the
transfer matrix $T_B(z)$.
The following is {\bf main result} of this section.
The free field realization of 
the boundary state $|B\rangle$
is given as following \cite{Kojima1}.
\begin{eqnarray}
|B\rangle=e^F|k,k\rangle.
\end{eqnarray}
Here we have set
\begin{eqnarray}
F&=&
-\frac{1}{2}\sum_{m>0}
\sum_{j=1}^{N-1}\sum_{k=1}^{N-1}
\frac{1}{m}\frac{[rm]_x}{[(r-1)m]_x}
I_{j,k}(m)\alpha_{-m}^j \alpha_{-m}^k
+\sum_{m>0}\sum_{j=1}^{N-1}\frac{1}{m}
D_j(m)\beta_{-m}^j,
\end{eqnarray}
where
\begin{eqnarray}
D_j(m)&=&
-\theta_m\left(\frac{[(N-j)m/2]_x[rm/2]_x^+
x^{\frac{(3j-N-1)m}{2}}}
{[(r-1)m/2]_x}\right)\nonumber\\
&&+\frac{x^{(j-1)m}[(-r+2\pi_{1,j}+2c-j+2)m]_x}{[(r-1)m]_x}
\nonumber\\
&&+\frac{[m]_x 
x^{(r-2c+2j-2)m}}{[(r-1)m]_x}
\left(\sum_{k=j+1}^{N-1}x^{-2m \pi_{1,k}}\right)\nonumber\\
&&+\frac{x^{(2j-N)m}[(r-2\pi_{1,N}-2c+N-1)m]_x}{[(r-1)m]_x},
\end{eqnarray}
and
\begin{eqnarray}
I_{j,k}(m)=\frac{[jm]_x[(N-k)m]_x}{[m]_x[Nm]_x}
=I_{k,j}(m)~~(1\leq j\leq k \leq N-1).
\end{eqnarray}
Here we have used
\begin{eqnarray}
~[a]_x^+=x^a+x^{-a},~~\theta_m(x)=
\left\{\begin{array}{cc}
x,& m:~{\rm even},\\
0,& m:~{\rm odd}.
\end{array}\right.\nonumber
\end{eqnarray}
Multiplying the type-II vertex operators 
$\Psi_\mu^{*(a,b)}(\xi)$
to the boundary state $|B\rangle$,
we get the diagonalization of
the boundary transfer matrix $T_B(z)$ on the space of state of the boundary
$U_{q,p}(\widehat{sl_N})$ face model.
It is thought that 
this method can be extended to the case of 
the elliptic quantum group $U_{q,p}(g)$
for affine Lie algebra $g$ \cite{JKOS}.

\section*{Acknowledgments}

This work is supported by the Grant-in-Aid for
Scientific Research {\bf C} (21540228)
from Japan Society for Promotion of Science.

\end{document}